\documentstyle[11pt,psfig]{article}

\newcommand{\coneone}{{c_{1\times 1}}}
\newcommand{\conetwo}{{c_{1\times 2}}}
\begin{document}

{\LARGE \bf
\begin{center}
Renormalization Group Flow of\\
SU(3) Gauge Theory
\end{center}
}

\bigskip


\begin{center}
{\large QCD-TARO Collaboration}
\end{center}

\begin{center}
 Ph.~de~Forcrand${}^a$,      M.~Garc{\'\i}a~P\'erez${}^b$,
 T.~Hashimoto${}^c$,         S.~Hioki${}^d$,
\\
 H.~Matsufuru${}^{e,f}$,    
 O.~Miyamura${}^e$,
 A.~Nakamura${}^g$,          I.-O.~Stamatescu${}^{f,h}$,
\\
 Y.~Tago${}^i$,              T. Takaishi${}^j$
 and  T.~Umeda${}^e$ 
\end{center}

\noindent
${}^a$ SCSC, ETH-Z\"urich, CH-8092 Z\"urich, Switzerland  \\
${}^b$ TH-Div., CERN, CH-1211, Geneve 23, Switzerland  \\
${}^c$ Dept. of Appl. Phys., Fac. of Engineering,
        Fukui Univ., Fukui 910-8507, Japan  \\
${}^d$ Dept. of Physics, Tezukayama Univ.,
        Nara 631-8501, Japan  \\
${}^e$ Dept. of Physics, Hiroshima Univ.,
        Higashi-Hiroshima 739-8526, Japan  \\
${}^f$ Inst. Theor. Physik, Univ. of Heidelberg,
        D-69120 Heidelberg, Germany  \\
${}^g$ Res. Inst. for Inform. Sci. and Education, Hiroshima Univ.,
        Higashi-Hiroshima  739-8521, Japan  \\
${}^h$ FEST, Schmeilweg 5, D-69118 Heidelberg, Germany \\
${}^i$ Dept. of Computat'l Sci., Kanazawa Univ.,
        Kanazawa 920-1192, Japan \\
${}^j$ Hiroshima University of Economics,
        Hiroshima 731-01, Japan  \\

\abstract{
We calculate numerically the renormalization group (RG) flow of 
lattice QCD in two-coupling space, 
$(\beta_{1\times 1},\beta_{1\times 2})$.
This is the first explicit calculation of the RG flow of
SU(3) gauge theory.
From the RG flow,
a renormalized trajectory (RT) is revealed.
Its behavior is consistent with the strong coupling
expansion near the high-temperature fixed point.
Actions with 
$(\beta_{1\times 1},\beta_{1\times 2})$
are studied;
the lattice spacing is evaluated by measuring the string tension
from the heavy quark potential.  
Recovery of the rotational symmetry is studied as a function of the ratio
$\beta_{1\times 2} / \beta_{1\times 1}$.
}

\newpage

\section{Introduction}

Since Wilson's first numerical RG analysis of SU(2) gauge theory \cite{Wilson},
there have been many Monte Carlo RG studies of non-perturbative
$\beta$-functions (see Ref.\cite{Gupta} and references therein).
In these analyses indirect information about the $\beta$ function,
such as $\Delta\beta$, has been obtained \cite{QCDTARO-MCRG}.
Recent progress of lattice techniques \cite{Okawa,HPW,Takaishi}
allows us to estimate directly the renormalization group (RG)
flow in a multi-coupling space\cite{patel_gupta}.
Here we will discuss  RG behavior of lattice QCD in the 
quenched approximation,
i.e., pure SU(3) gauge theory.  

New blocked actions $S'$ as a function of blocked link variables $U^B$'s 
are constructed from the original $S(U)$ as
 
\begin{equation}
e^{-S'(U^{B})} = \int e^{-S(U)} \delta(U^{B}-P(U))  d U ,
\end{equation}

\noindent
where $P$ defines the blocking transformation.
$S$ and $S'$ are different points in the infinite coupling 
space: after the blocking transformation $S$ moves into $S'$.
Only fixed points are left invariant. In this process
the lattice spacing, i.e., the cut-off of the theory, is
changed, and this RG flow characterizes the theory.

Corresponding to each blocking scheme $P$ there is a renormalized 
trajectory (RT), which starts at the ultra-violet fixed point.
More than ten years ago, Iwasaki estimated a RT by matching Wilson
loops (obtained in a certain approximation), and proposed to use it as
an improved action \cite{Iwasaki0}.
Recently Hasenfratz and Niedermayer have reminded us  that
any point on  the RT is a ``perfect action'' \cite{perfect},
since there the long distance behavior must be same as at the
fixed point. 
Therefore if we find a RT, we have a perfect action.  
Even if it is an approximate one, it may serve as an improved
action.

In this paper we will determine the RG flow
in the following approximation; 
the coupling space is restricted to two parameters
and the action is assumed to have the following form,

\begin{eqnarray}
S & = & \beta \{ {\coneone} \sum_{plaq} 
          (1 - {1 \over 3}{\rm Re Tr} U_{plaq}) \nonumber \\
  & + & {\conetwo} \sum_{rect} 
            (1 - {1 \over 3}{\rm Re Tr} U_{rect}) \} \nonumber \\
  & = & - \frac{1}{2} a^4 ({\coneone}+8 {\conetwo}) \beta 
  \sum {\rm Tr} F_{\mu\nu}(x)^2
      + O(a^6) .
\label{eq:action}
\end{eqnarray}

\noindent
Here $a$ is the lattice spacing and $U_{plaq}$ and $U_{rect}$ correspond to
$1\times 1$ and $1\times 2$ loops respectively.
We impose the condition,
${\coneone}+8 {\conetwo} = 1$.

\section{Technique to determine the RG flow}

We adopt Swendsen's factor-two blocking scheme 
\cite{Swendsen}.  The blocked link variable is constructed as:

\begin{equation}
Q_{\mu}(x) = U_{\mu}(x) U_{\mu}(x+{\mu}) 
       + \frac{1}{2} \sum_{\nu \ne \mu} U_{\nu}(x) U_{\mu}(x+{\nu}) 
                    U_{\mu}(x+{\nu}+{\mu})U_{\nu}^{\dag}(x+2{\mu}).    
\end{equation}
\noindent
Projecting  $Q_{\mu}(x) $  onto SU(3), we get a new blocked variable $U'$
as
$ \mbox{max} \{ Re~{\rm Tr}~Q_{\mu}(x) U_{\mu}^{\dag}(x) \} $.

From the original configurations, $\{U\}$,  generated by 
the action $S$
with parameters $({\coneone},{\conetwo})$ in Eq.(\ref{eq:action}),
we obtain the blocked ones,
$\{U'\}$, after blocking.
If the blocked configurations, $\{U'\}$, can be considered 
as generated
by an action $S'$ with parameters $({\coneone}',{\conetwo}')$ 
in Eq.(\ref{eq:action}),
then 
$ ({\coneone},{\conetwo}) \rightarrow ({\coneone}',{\conetwo}')
$
can be regarded as the coupling flow associated with this blocking.

For the determination of the effective action $S'$, we use a 
Schwinger-Dyson method \cite{Okawa}.
This method is based on the following identity. 
For a link $U_l$, consider the quantity;

\begin{equation}
<{\rm Im}~{\rm Tr} (\lambda^c U_l G_l^{\alpha})> 
=  \int DU \ {\rm Im}~{\rm Tr}
(\lambda^c U_l G_l^{\alpha})
e^{-S}/Z,
\label{SD0}
\end{equation}
\noindent
where $\lambda^c$ stands for Gell-Mann matrices. 
$G_l$ is a sum of ``staples'' $G_l^{\gamma}$ for the link $l$,
$G_l = \sum_{\gamma} {(\beta_{\gamma} / 6)} G_l^{\gamma}$.
The action $S$ is assumed to  have the form  
$\sum_{l}~Re~{\rm Tr} U_l G_l$.
For the present analysis, $\gamma$ corresponds to a plaquette
and a rectangle.
Eq.(\ref{SD0}) should be invariant under the change of variables
$U_l \rightarrow (1+i\epsilon\lambda^{c})U_l$.
Setting the terms linear
in $\epsilon$ to zero, we get the identity,

\vspace*{-5mm}
\begin{equation}
  \int DU
[ \ {\rm Re}~{\rm Tr}((\lambda^c)^2 U_l G_l^{\alpha}) \\
+ {\rm Im}~{\rm Tr}(\lambda^c U_l G_l^{\alpha}) {\rm Im} {\rm Tr}(\lambda^c U_l G_l)] \nonumber \\
 e^{-S} = 0. \ \ 
\label{SD1}
\end{equation}

Summing over $c$ in the formula (\ref{SD1}), we obtain the Schwinger-Dyson
equation,
\begin{eqnarray}
{8 \over 3}{\rm Re}<{\rm Tr}(U_l G_l^{\alpha})> = 
 \sum_{\gamma} {\beta_{\gamma} \over 6}     
&\{& -  {\rm Re} <{\rm Tr}(U_lG_l^{\alpha}U_lG_l^{\gamma})>  
\label{eq:SD}\\ 
 +  {\rm Re} <{\rm Tr}(G_l^{\alpha}(G_l^{\gamma})^{\dagger})> 
 &-& {2 \over 3} <{\rm Im~Tr}(U_lG_l^{\alpha}) 
{\rm Im~Tr}(U_lG_l^{\gamma})>\}. \nonumber 
\end{eqnarray}
\noindent
Here we used $\sum_{c=1}^8 {\rm Tr}(\lambda^c A) 
{\rm Tr}(\lambda^c B) = 2 {\rm Tr} AB - \frac{2}{3} {\rm Tr} A {\rm Tr} B$.
We apply this equation to the blocked configurations, and
calculate the expectation values $<\cdots>$ on both sides. Now
Eq.(\ref{eq:SD}) may be considered as a set of linear equations with
$\beta_\gamma$'s as unknowns. 

In general we may use other quantities instead of $G^{\alpha}$,
but the number of equations becomes equal to the number
of unknowns if we take $G^{\alpha}$ appearing in our action.
In comparison with the demon method \cite{demon} , larger loops such as 
${\rm Tr}(G_l^{\alpha}(G_l^{\gamma})^{\dagger})$
are involved in the Schwinger-Dyson equation.

\section{Results of RG flow}

We have used lattices of size  $8^4$ and $16^4$.
About 2000 configurations separated by
every 10 sweeps are used to determine each parameter set
$(\beta_{1\times1},\beta_{1\times2}) = \beta (\coneone,\conetwo)$.

\begin{figure}[hbt]
\begin{center}
\psfig{file=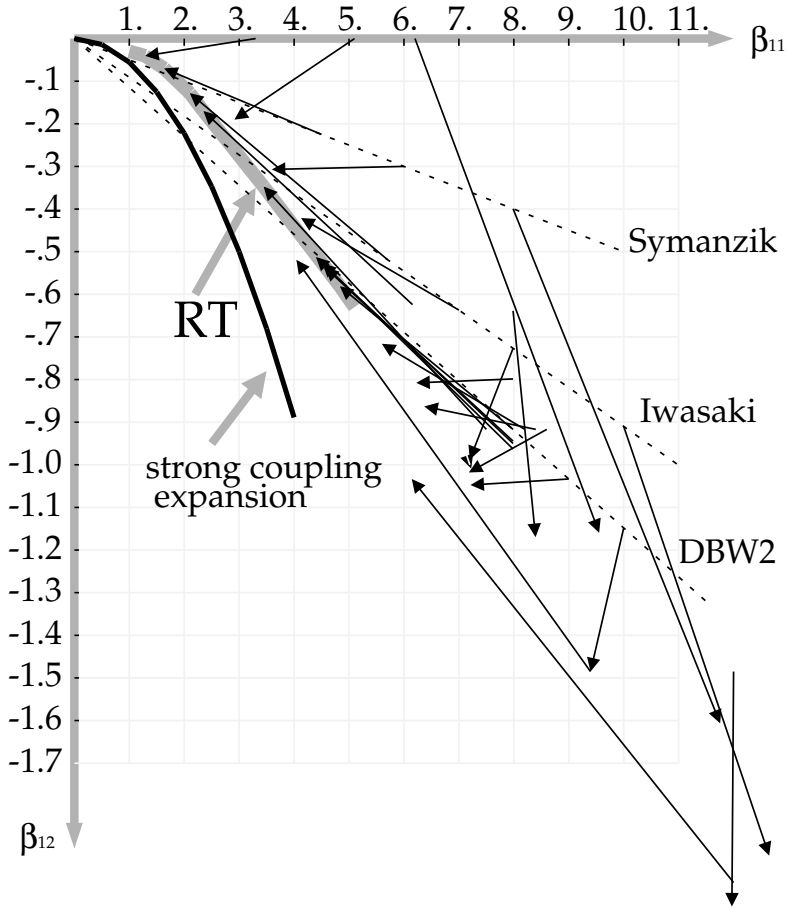,width=100mm,height=120mm,bbllx=0bp,bblly=116bp,bburx=445bp,bbury=645bp}
\caption {Renormalization group flow for QCD in two coupling space}
\label{fig:flow}
\end{center}
\end{figure}

In Fig.\ref{fig:flow}, the result of the coupling flow is shown. 
Arrows denote the measured coupling flow.
We perform a Monte Carlo simulation with $({\coneone},{\conetwo})$ 
(the starting point of an arrow),
and then block the lattice. 
From the blocked links we estimate a blocked action 
(the end point) through the formula (\ref{eq:SD}) above,
by assuming that the blocked action has only two terms, 
plaquette (${\coneone}'$)  and rectangle (${\conetwo}'$).
It is remarkable that the arrows show a flow of simple structure.
They flow into a trajectory.  This is a strong evidence that the
RT corresponding to this blocking scheme can be seen even in the 
two-coupling space.

At the strong coupling region, the RT does not look like a linear curve
but like a parabolic curve.  
This is completely consistent with the strong coupling expansion. 
In the strong coupling regime, $\beta$ is small, and the Boltzmann weight 
$e^{-\beta S}$ can be expanded as a polynomial in $\beta$. 
If we include the $1 \times 2$ rectangle in the action
in addition to the $1 \times 1$ plaquette,
then the first few loops can easily be evaluated in the regime where
$\beta_{1\times1}, \beta_{1\times2} \ll 1$. 
In particular, the plaquette and the rectangle give: 
$\langle W_{11} \rangle \approx \frac{\beta_{1\times1}}{18}$ , and 
$\langle W_{12} \rangle \approx (\frac{\beta_{1\times1}}{18})^2 
+ \frac{\beta_{1\times2}}{18}$. 
The string tension can then be evaluated at strong coupling as: 
\begin{equation}
\sigma a^2 \approx 
- \log \frac{\langle W_{12} \rangle}{\langle W_{11} \rangle}
= - \log \left(\frac{\beta_{1\times1}}{18} 
+ \frac{\beta_{1\times2}}{\beta_{1\times1}} \right). 
\end{equation}
Consider now the strong coupling limit 
$\beta_{1\times1}, \beta_{1\times2} \rightarrow 0$.
The left-hand side should go to $+\infty$, indicating
that the lattice spacing $a$ diverges in the strong coupling limit
\cite{Creutz}.

What is remarkable in the expression for the string tension is that 
it does not diverge unless 
$\frac{\beta_{1\times1}}{18} 
+ \frac{\beta_{1\times2}}{\beta_{1\times1}} \rightarrow 0$.
In other words, as the renormalized trajectory approaches the high-temperature
fixed point (HTFP) ($\beta_{1\times1}=\beta_{1\times2}=0$), 
it must approach the parabola 
\begin{equation}
\beta_{1\times2} = - \frac{1}{18} \beta_{1\times1}^2.
\end{equation}


As one goes away from the HTFP along the RT, the lattice spacing shrinks and
larger loops must be considered to evaluate the string tension. A
straightforward evaluation of the Creutz ratio $\chi_{22}$ already shows
that the parabola gets distorted ``upwards'' as one considers points on
the RT further and furtheraway from the HTFP. 

On the other hand, in the vicinity of the Ultra-Violet Fixed Point (UVFP),
where the lattice spacing is very small, the usual Symanzik improvement
(tree-level, tadpole, one-loop or non-perturbative) should be effective at removing
${\cal O}(a^2)$ cutoff effects. So one expects, as one approaches the UVFP,
that the ratio $-\beta_{1\times2} / \beta_{1\times1}$ will approach the
Symanzik value 0.05. Our MCRG study shows that this ratio is approached from 
above, as seen also when implementing the Symanzik scheme beyond tree-level.
Here we exhibit the smooth connection from the UVFP to the HTFP in the 
two-coupling plane.

\section{Actions in two-coupling space}

The analysis in the previous section
indicates that along the RT the ratio 
$-c_{1\times 2}/c_{1\times 1}$  should increase
with the lattice spacing, at least until the lattice becomes
very coarse. Here we study the rotational symmetry of the heavy quark
potential of several
approximations to the RT, characterized by a fixed value of 
$c_{1\times 2}/c_{1\times 1}$.

We parameterize the two-dimensional plane by lines, 
${\coneone}+8{\conetwo}=1$, with ${\conetwo}$ as a parameter; 
${\conetwo}=0$ corresponds to the standard Wilson action, 
${\conetwo}=-1/12$ to Symanzik's \cite{Symanzik}
(tree level with no tadpole improvement),
${\conetwo}=-0.331$ to Iwasaki's \cite{Iwasaki0}.
We include one more steeper line with ${\conetwo}=-1.4088$
which we call DBW2. This line goes through  the point 
obtained after double blocking from a lattice with Wilson 
action ($\beta =6.3$, $32^3\times64 
\rightarrow \beta_{1 \times 1} \approx 7.986,
\beta_{1 \times 2} \approx -0.9169$) in 2-dimensional coupling 
space by the canonical demon method \cite{Forcrand-Takaishi}.
 
First let us determine the lattice spacing.
We measure the string tension, $\sigma_{phys} a^2$
from the heavy quark potential.  
The heavy quark potential is obtained from Wilson loops $W(R,T)$
calculated from smeared links. The
smearing technique proposed in \cite{Smear} reduces contaminations 
from excited states to a very large extent. 

To determine the heavy quark potential $V(R)$,
links are first smeared along the spatial directions; 
$W(R,T)$ is then computed in terms of the smeared links.
The Wilson loops are expected to behave as
\begin{equation}
W(R,T) = C(R)\exp{[-V(R)T]} + \sum C_i(R)\exp{[-V_i(R)T]}
\label{eq1}
\end{equation}
where $C(R)$ is the overlap function and $V(R)$ is the heavy quark
potential, the second term is the contribution from excited states.
We obtain $V(R)$ by applying  least square fitting 
to the first term in Eq.(\ref{eq1}) as a function of $T$ for each $R$. 
The fitting range of $T$ is chosen for each $R$ such that the
effective mass plots in $T$ become approximately constant
neglecting excited states in the fitting. 
The number of smearing steps is determined so that the overlap function $C(R)$ 
is closest to one for each $R$.

Next we extract the string tension from $V(R)$.
We take the following Ansatz,
\begin{equation}
V(R)= A + \frac{\alpha}{R} + \sigma R \label{eq2}
\end{equation}
The string tension, $\sigma=\sigma_{phys}a^2$, is the coefficient 
of the linear term in Eq.(\ref{eq2}).
We fit the heavy quark potential to Eq.(\ref{eq2})
in the range where $R a$ corresponds approximately to
physical distances about $0.4 fm \sim 1.5 fm$.
Statistical errors are estimated by the jackknife method 
with bin size one.
We set $\sqrt \sigma_{phys} =420 MeV$ and fix the scale $a$.
Results are summarized in Table \ref{beta-a}.

Simulations are performed on  a lattice of  size  
$12^3 \times 24$ . Thermalization is 5000 sweeps,
while the interval between Wilson loop measurements is 500 sweeps.
The number of configurations used is 100.

\begin{table}[h]
\begin{center}
\begin{tabular}{c|cccc}
\hline
\hline
action&$\beta$ &$\sigma $&$a[\mbox{fm}]$ &$\Delta$\\
\hline
Wilson action          & 5.45 & 0.463(45) & 0.320(16) & 0.082(16) \\
$\coneone=1$           & 5.55 & 0.2840(46)& 0.2504(20)& 0.049(10) \\
$\conetwo=0$           & 5.65 & 0.1936(72)& 0.2067(38)& 0.0254(20)\\
\hline
Symanzik action (tree) & 3.70 & 0.4666(89)& 0.3209(31)& 0.056(13)\\
$\coneone=1-8\conetwo$ & 3.90 & 0.2634(41)& 0.2411(19)& 0.0317(53)\\
$\conetwo=-1/12$       & 4.10 & 0.1377(48)& 0.1743(30)& 0.0123(12)\\
\hline
Iwasaki action         & 1.90 & 0.6363(72)& 0.3748(21)& 0.0361(65)\\
$\conetwo=-0.331$      & 2.00 & 0.465(13) & 0.3204(45)& 0.0070(45)\\
                       & 2.10 & 0.3496(79)& 0.2778(31)& 0.0114(27)\\
                       & 2.20 & 0.2031(75)& 0.2117(39)& 0.0057(28)\\
                       & 2.30 & 0.1333(63)& 0.1715(41)& 0.0071(28)\\
\hline
DBW2 action            & 0.62 & 0.687(34) & 0.3894(96)& 0.0147(17)\\
$\conetwo=-1.4088$     & 0.65 & 0.512(18) & 0.3362(59)& 0.0224(50)\\
                       & 0.70 & 0.362(13) & 0.2827(51)& 0.0062(24)\\
                       & 0.75 & 0.2489(71)& 0.2344(33)& 0.0054(16)\\
                       & 0.80 & 0.1631(55)& 0.1897(12)& 0.0078(30)\\
\hline
\hline
\end{tabular}
\end{center}
\caption{String tensions and the lattice spacing determined from them} 
\label{beta-a}
\end{table}

If the action lies on the RT, long range quantities behave in the same way
as in the continuum.  Therefore, even if the lattice is coarse, 
the rotational symmetry is expected to be recovered.
Now we investigate the symmetry breaking in the $(\coneone,\conetwo)$
plane.

We construct a quantity which measures the amount of rotational symmetry
breaking;
\begin{equation}
\Delta^2 \equiv 
\sum_{\mbox{off-axis}}
\frac{[V(R)-V_{\mbox{on}}(R)]^2}{V(R)^2\delta V(R)^2} 
\left(
\sum_{\mbox{off-axis}}\frac{1}{\delta V(R)^2}
\right)^{-1}.
\end{equation}
where $V_{\mbox{on}}(R)$ is a fit to Eq.(\ref{eq2})
determined from on-axis data only.
Here ``on-axis data'' are those on points
along a lattice axis.
$\delta V(R)$ is the statistical error of $V(R)$.
We employ 
$\Delta$ as a measure of the difference between on-axis and off-axis 
potential, and therefore a measure of the rotational symmetry breaking. 
We calculate this quantity with three improved actions at various
lattice spacings, i.e., various $\beta$'s.

\begin{figure}[h]
\begin{center}
\psfig{file=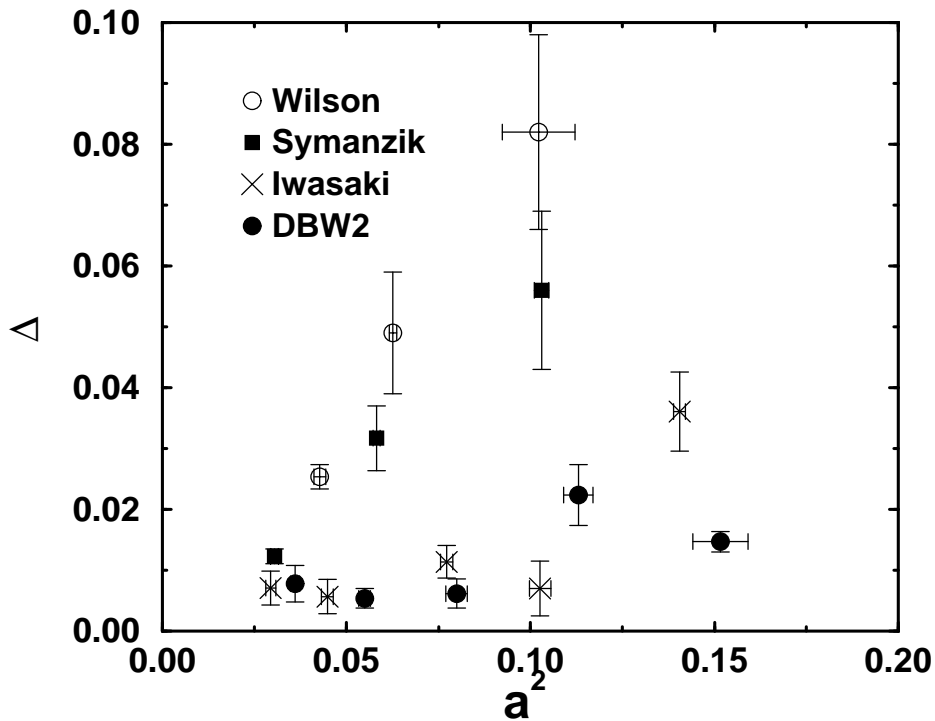,width=140mm,height=100mm,bbllx=-50bp,bblly=21bp,bburx=527bp,bbury=436bp}
\caption {Rotational symmetry breaking}
\label{fig1}
\end{center}
\end{figure}

Results  are shown in Table \ref{beta-a} and Fig.\ref{fig1}.
For small $\conetwo$, like Wilson or tree-level Symanzik actions with
no tadpole improvement, $\Delta$ is proportional to $a^2$. 
As $|\conetwo|$ increases, the rotational symmetry is better
recovered. To the present accuracy and in the range of couplings studied, 
the breaking of the rotational symmetry can be already ignored
at the level of Iwasaki's  action.

\section{Concluding remarks}

\begin{figure}[h]
\begin{center}
\psfig{file=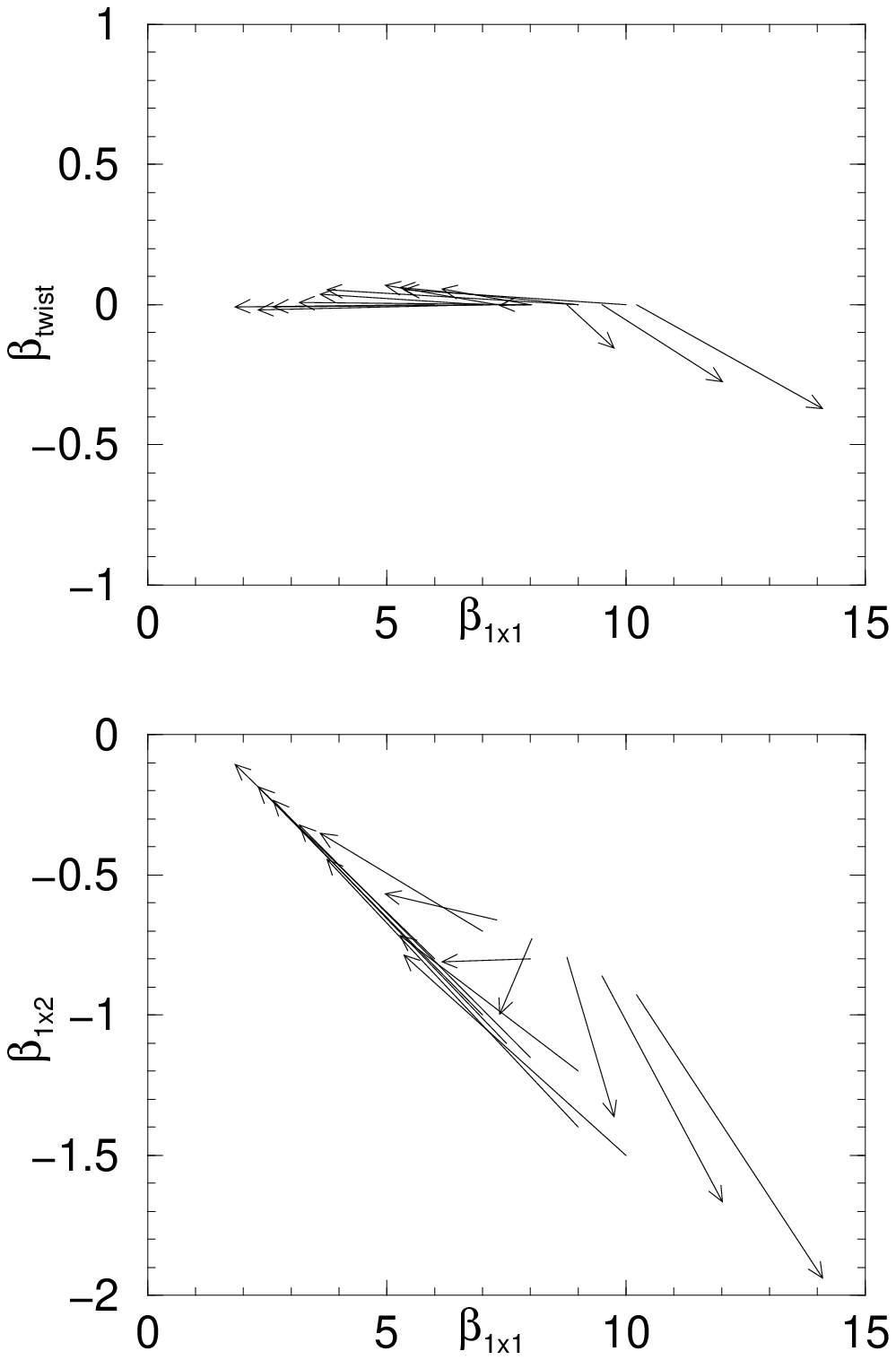,width=50mm,height=100mm,bbllx=35bp,bblly=26bp,bburx=240bp,bbury=466bp}
\caption{RG flow in ($\beta_{1\times1},\beta_{1\times2}$) plane, and
($\beta_{1\times1},\beta_{twist}$) plane.}
\label{fig-rg2}
\end{center}
\end{figure}

In this paper, we calculate numerically the RG flow in two-coupling space
$(\coneone,\conetwo)$, as the first step to study the RG flow. 
Although it is very remarkable that the flow
structure can be already observed, the results may have ``truncation'' errors. 
After a blocking transformation, the action has in general infinitely
many couplings. Therefore end points of the arrows in Fig.\ref{fig:flow}
generally deviate from the two-dimensional plane. 
To estimate this effect, we calculate the behavior of arrows, whose starting
points are near the RT on the two-dimensional plane,  
in three-dimensional space: 
$(\beta_{1 \times 1},\beta_{1 \times 2},\beta_{twist})$,
see Fig.\ref{fig-rg2}. 
Deviations from the $(\beta_{1 \times 1},\beta_{1 \times 2})$ plane are
small especially near the RT.  This is expected also from the strong
coupling expansion: because it takes 3 plaquettes to cover the ``twist''
loop, its effective coupling will behave like $\beta_{1 \times 1}^3$ 
near the HTFP. 
 
It is very encouraging that here we have seen a clear flow behavior
which strongly indicates the existence of a renormalized trajectory
in the vicinity of the 2-coupling plane.
We now plan to continue the Monte
Carlo RG calculation 
(i) at higher $\beta$'s, in order to obtain the non-perturbative
QCD $\beta$ function, and (ii) in three- or four-coupling space
including the chair and twist operators, 
to estimate truncation effects and evaluate the reliability
of the results obtained here.

All simulations have been done on CRAY J90 
at Information Processing Center,
Hiroshima University,
SX-4 at RCNP, Osaka university 
and on VPP500 at KEK (National Laboratory for High Energy Physics). 
H.M. would like to thank the Japan Society for the
Promotion of Science for Young Scientists for 
financial support.

\end{document}